\documentstyle[preprint,prl,aps]{revtex}

\def\bec{\begin{center}}
\def\eec{\end{center}}

\def\beq{\begin{equation}}
\def\eeq{\end{equation}}

\renewcommand{\frac}[2]{{{\displaystyle #1}\over{\displaystyle #2}}}

\def\ltap{\ \raisebox{-.4ex}{\rlap{$\sim$}} \raisebox{.4ex}{$<$}\ }

\def\ltap{\lsim}  

\def\ltap{\ \raisebox{-.4ex}{\rlap{$\sim$}} \raisebox{.4ex}{$<$}\ }

\begin{document}
\sloppy

\vspace{0.5cm}

\draft
\begin{titlepage}
\newpage
\preprint{\vbox{\baselineskip 10pt{
\hbox{Ref. SISSA 32/2000/EP}
\hbox{March, 2000}
\hbox{hep -- ph/0004151}
}}}
\vskip -0.4cm
\title{\bf The Super - Kamiokande Day - Night Effect Data and \\
       the MSW Solutions of the Solar Neutrino Problem}

\author{M. Maris$^{a}$ and S.T. Petcov $^{b,c)}$\footnote{Also at:
Institute of Nuclear Research and Nuclear Energy, Bulgarian Academy
of Sciences, BG--1784 Sofia, Bulgaria.}}
\address{a) Osservatorio Astronomico di Trieste, I-34113 Trieste, Italy\\
\vskip -0.4cm
b) Scuola Internazionale Superiore di Studi Avanzati, I-34014 Trieste, Italy\\
\vskip -0.4cm
c) Istituto Nazionale di Fizica Nucleare, Sezione di Trieste, I-34014
Trieste, Italy}
\vskip -0.4cm
\maketitle
\begin{abstract}
\baselineskip 16pt
\tightenlines 

The current Super-Kamiokande data 
on the D-N asymmetry between the
the day event rate and the
{\it Night} ({\it Mantle}) and {\it Core} event rates, produced
by solar neutrinos which respectively cross the Earth along
any trajectory (cross the Earth mantle but do not cross the core),
and cross the Earth core 
before reaching the detector, 
imply rather stringent constraints
on the MSW small mixing angle (SMA)
$\nu_e \rightarrow \nu_{\mu(\tau)}$
solution of the solar neutrino problem.
A simplified analysis shows, in particular,
that a substantial subregion of the SMA solution region
is disfavored by these data. 
The {\it Core} D-N asymmetry data alone allow 
to rule out at 99.7\% C.L.
a part of this subregion.
The constraints on the
MSW large mixing angle and 
LOW $\nu_e \rightarrow \nu_{\mu(\tau)}$
solutions as well as on the 
MSW $\nu_e \rightarrow \nu_{s}$ 
solution, following from the 
data on the {\it Mantle}, {\it Night}  
and {\it Core} D-N asymmetries are also discussed.

\end{abstract}

\end{titlepage}

\newpage
\tightenlines

\leftline{\bf I. Introduction: The 
MSW Solutions of the Solar Neutrino Problem}
{\bf and the Day-Night Effect}
\vskip 0.3cm

The hypothesis of MSW transitions 
of solar neutrinos continues 
to provide a viable solution of the 
solar neutrino problem 
\cite{Cl,K,SAGE,GALLEX,SK99,BP98,Fogli99,CGG99,PK99}. 
  The current (mean event rate) 
solar neutrino data admit three types of 
MSW $\nu_e \rightarrow \nu_{\mu (\tau)}$ 
transition solutions:
the well-known small mixing angle 
(SMA) non-adiabatic
and large mixing angle (LMA) adiabatic 
(see, e.g., \cite{KPMSW93,HataL94}) and 
the so-called ``LOW'' solution 
(very recent analyses can be found 
in, e.g., \cite{SK99,Fogli99,CGG99,PK99}).
While the SMA and LMA solutions have been shown 
to be rather stable with respect to variations 
in the values of the various physical quantities which 
enter into the calculations (the fluxes of 
$^8$B and $^7$Be neutrinos, 
nuclear reaction cross-sections, etc.),
and of the data utilized in the analyses,
the LOW solution is of the ``borderline'' 
type: its existence even at 99\% C.L. 
is not stable with respect to
relatively small changes in the data 
and/or in the 
relevant theoretical predictions 
(see, e.g., \cite{Fogli99} and 
the references quoted therein.).
  
    To the three solutions there correspond
(at a given C.L.)
three distinct regions in the plane of values 
of the two parameters, 
$\Delta m^2$ and $\sin^22\theta$, 
characterizing the transitions.
One finds \cite{SK99,Fogli99,CGG99,PK99} using the 
standard solar model predictions 
\cite{BP98}  for the solar neutrino fluxes 
($^8$B, $^7$Be, $pp$, etc.) 
that at 99\% C.L. the SMA MSW solution requires
values in the intervals 
$4.0 \times 10^{-6}~{\rm eV^2} \ltap \Delta m^2 \ltap  
10.0\times 10^{-6}~{\rm eV^2}$,
$1.3 \times 10^{-3} \ltap \sin^22\theta 
\ltap 1.0\times 10^{-2}$,
the LMA solutions is realized for
$\Delta m^2$ and $\sin^22\theta$
from the region
$7.0\times 10^{-6}~{\rm eV^2} \ltap \Delta m^2 \ltap  
2.0\times 10^{-4}~{\rm eV^2}$,
$0.50 \ltap \sin^22\theta 
\ltap 1.0$, and the LOW solution 
lies approximately in the region
$0.4\times 10^{-7}~{\rm eV^2} \ltap \Delta m^2 \ltap  
1.5\times 10^{-7}~{\rm eV^2}$,
$0.80 \ltap \sin^22\theta 
\ltap 1.0$. 
The SMA and LMA solution regions expand
in the direction of smaller values of 
$\sin^22\theta$ up to
 $\sim 0.6\times 10^{-3}$ and 
to $\sim 0.3$, respectively, 
if one adopts a more {\it conservative} approach 
in analyzing the data in terms of the MSW effect 
and treats the $^8$B neutrino flux as a free parameter
in the analysis (see, e.g., \cite{PK99}). 

  A unique testable prediction of the 
MSW solutions of the solar 
neutrino problem is the day-night (D-N) effect -
a difference between the 
solar neutrino event rates
during the day and during the night,
caused by the additional transitions of the solar
neutrinos taking place at night while the neutrinos 
cross the Earth on the way to the detector 
(see, e.g., \cite{HataL94,DNold} and the references quoted therein).
The experimental observation of a non-zero  D-N asymmetry
\begin{equation}
A^{N}_{D-N} \equiv \frac{R_{N} - R_{D}}{(R_N + R_D)/2},
\end{equation}
\noindent where $R_N$ and $R_D$ are, e.g., the  
one year averaged event rates in a given detector 
respectively during the night and the day, 
would be a very strong evidence in favor 
(if not a proof) of the MSW solution
of the solar neutrino problem. 

  Extensive predictions 
for the magnitude of the D-N effect
for the Super-Kamiokande detector have been obtained 
in \cite{SK97I,SK97II,SK98III,LisiM97,BK97}.
Earlier results have been derived in \cite{HataL94,DNold}.
High precision calculations of
the D-N asymmetry in the 
one year averaged recoil-e$^{-}$ spectrum and  
in the energy-integrated event rates were performed
for three event samples, 
{\it Night}, {\it Core} and {\it Mantle},
in \cite{SK97I,SK97II,SK98III}.
The night fractions of these event samples 
are due to neutrinos which respectively cross 
the Earth along any trajectory, 
cross the Earth core, and
cross only the Earth mantle (but not the core),
on the way to the detector. 
The measurement of the D-N asymmetry in the 
{\it Core} sample was found to be
of particular importance \cite{SK97I,SK97II} 
because of the 
strong enhancement of the asymmetry,
caused by a constructive interference  
between the amplitudes of the
neutrino transitions in the Earth 
mantle and in the Earth core 
\cite{SPPLB43498}.
The effect differs from the MSW one \cite{SPPLB43498}.
The  {\it mantle-core enhancement effect} 
is caused by the existence (for a given neutrino trajectory
through the Earth core) of 
{\it points of resonance-like 
total neutrino conversion}
in the corresponding space 
of neutrino oscillation 
parameters \cite{ChPet991,ChPet992}. 
The location of these points determines the regions
where the relevant probability of transitions
in the Earth of the Earth-core-crossing solar neutrinos
is large 
\footnote{ Being a
constructive interference effect between 
the amplitudes of neutrino transitions 
in the mantle and in the core,
this is not just ``core enhancement'' effect,
but rather {\it mantle-core enhancement} effect.}
\cite{ChPet992}.
At small mixing angles 
and in the case of 
$\nu_e \rightarrow \nu_{\mu (\tau)}$ 
transitions
the predicted D-N asymmetry in the 
{\it Core} sample of the Super-Kamiokande
event rate data was shown \cite{SK97II} to 
be much bigger due to the 
{\it mantle-core enhancement} effect 
\footnote{The term ``neutrino oscillation 
length resonance'' (NOLR) was used in 
\cite{SPPLB43498}, in particular,
to denote the enhancement in this case.}
-  by a factor of up to $\sim 6$,
than the asymmetry in the {\it Night} sample.
The asymmetry in the {\it Mantle} 
sample was found to be smaller than the 
asymmetry in the {\it Night} sample. 
On the basis of these results it was 
concluded in \cite{SK97II} that 
it can be possible to test a substantial part of the
MSW $\nu_e \rightarrow \nu_{\mu (\tau)}$ SMA 
solution region in the $\Delta m^2 - \sin^22\theta$
plane by  performing selective, i.e., 
{\it Core} and {\it Night} (or {\it Mantle}) 
D-N asymmetry measurements.

  The current Super-Kamiokande data \cite{SK99} shows a    
D-N asymmetry in the {\it Night} sample, which is 
different from zero at 1.9 s.d. level: 
\begin{equation}
A^{N}_{D-N} = 0.065 \pm 0.031~(stat.) \pm 0.013~(syst.).
\end{equation}
\noindent These data allow to probe 
only a relatively small subregion of 
the SMA ``conservative'' solution region: 
the predicted asymmetry is too small 
(see, e.g., \cite{SK97II,LisiM97,BK97}).
However, the Super-Kamiokande night data is given in
5 bins and 80\% of the events in the bin N5  
are due to Earth-core-crossing solar neutrinos \cite{SK99},
while the remaining 20\% are produce by neutrinos which cross 
only the Earth mantle. Since the predicted D-N asymmetry
in the {\it Mantle} sample is practically negligible 
in the case of the MSW SMA solution of interest \cite{SK97II},
we have for the D-N asymmetry measured using the  
night N5 bin data: $A^{N5}_{D-N} \cong 0.8A^{C}_{D-N}$,
$A^{C}_{D-N}$ being the asymmetry in the 
{\it Core} sample. The data on $A^{N5}_{D-N}$ \cite{SK99}
permitted to exclude a part of the MSW SMA solution region 
located in the area
$\sin^22\theta \cong (0.007 - 0.01)$,
$\Delta m^2 \cong (0.5 - 1.0)\times 10^{-5}~{\rm eV^2}$.
It should be obvious from the above discussion that
the measurement of the {\it Core} asymmetry
$A^{C}_{D-N}$, as suggested in \cite{SK97II},
will provide a more effective  test
of the the MSW SMA solution than the
measurement of $A^{N5}_{D-N}$.

 Recently the Super-Kamiokande collaboration 
for the first time published data on 
the {\it Core} D-N asymmetry
$A^{C}_{D-N}$ \cite{SK00}: 
    
\begin{equation}
A^{C}_{D-N} = -0.0175 \pm 0.0622~(stat.) \pm 0.013~(syst.). 
\end{equation} 

\noindent The experimental value of the {\it Mantle} asymmetry  
can also be deduced from the data \cite{SK00}:

\begin{equation}
A^{M}_{D-N} = 0.0769 \pm 0.034~(stat.) \pm 0.013~(syst.). 
\end{equation}

  In the present article we use the Super-Kamiokande results 
on the D - N effect, eqs. (2), (3) and (4), to derive constraints 
on the MSW $\nu_e \rightarrow \nu_{\mu (\tau)}$ transition 
solutions of the solar neutrino problem. 
We show below, in particular, that, as has 
been suggested in \cite{SK97II}, the data on the 
{\it Core} and {\it Night} or {\it Mantle} 
D - N asymmetries allow to perform a very effective
test of the MSW SMA solution. 
We also obtain 
constraints on the
MSW LMA and 
LOW $\nu_e \rightarrow \nu_{\mu(\tau)}$
solutions as well as on the 
MSW $\nu_e \rightarrow \nu_{s}$ 
solution, following from the 
Super-Kamiokande
data on the {\it Mantle}, {\it Night} 
and {\it Core} D-N asymmetries.
\newpage

\vskip 0.3cm
\leftline{\bf II. Constraints on the MSW Solutions from
the Super-Kamiokande Data}
{\bf on the {\it Core} and {\it Mantle} or
{\it Night} Day-Night Asymmetries}
\vskip 0.3cm 
  
  We use the high precision methods of calculation
of the energy-integrated one year average
{\it Core}, {\it Night} and {\it Mantle} 
D-N asymmetries
developed for our earlier 
studies of the D-N effect for the Super-Kamiokande detector, 
which are described in detail in \cite{SK97I,SK97II,SK98III}.
The cross section of the
$\nu_e - e^-$ elastic scattering reaction 
was taken from \cite{ESCross}.
We used in our calculations 
the $^8$B neutrino spectrum  
derived in 
\cite{B(8)nuSpectrum}. 
The probability
of survival of the solar $\nu_e$ when they travel 
in the Sun and further to the surface of the Earth,
$\bar{P}_{\odot}(\nu_e \rightarrow \nu_e)$,
was computed on the basis of the 
analytic expression obtained in \cite{SPMSWExp}
and using the method developed in
\cite{KPMSW88}. In the calculation of 
$\bar{P}_{\odot}(\nu_e \rightarrow \nu_e)$
we have utilized (as in \cite{SK97I,SK97II,SK98III})  
the density profile 
of the Sun and the $^8$B neutrino production 
distribution in the Sun, predicted in \cite{BP95}.
These predictions were updated in \cite{BP98},
but a detailed test study showed 
\cite{Maris99} (see also \cite{BK97})
that using the results of \cite{BP98} instead 
of those in \cite{BP95} leads to a change of 
the survival probability by less than 1\%
for any set of values of the parameters
$\Delta m^2$ and $\sin^22\theta$, 
relevant for the calculation of the D-N effect. 
As in \cite{SK97I,SK97II,SK98III},
the Earth matter effects were calculated 
using the Stacey model from 1977 \cite{Stacey77}
for the Earth density distribution. 
The latter practically coincides with 
that predicted by the more recent Earth model 
\cite{PREM81}. 
Our results are obtained for 
the standard values of
the electron fraction number in the Earth mantle and
the Earth core, 
$Y_e^{man} = 0.49$ and $Y_{e}^{c} = 0.467$, which reflect
the chemical composition of 
the two major Earth structures
(see, e.g, \cite{YeEarth,SK97II}). 

 In our simplified analysis
we use the Super-Kamiokande data on the {\it Core} and 
{\it Mantle} or {\it Night} D-N asymmetries,
treating  the {\it Core} and 
{\it Mantle}, and the {\it Core} and 
{\it Night} asymmetries as two pairs of 
independent observables.
While this is certainly justified in the case of the 
{\it Core} and {\it Mantle} asymmetries,
the treatment of the
{\it Core} and {\it Night} 
asymmetries neglects
the possible (weak) correlation between the values 
of the two asymmetries.
Note, however, that i) 
since such a correlation 
does not exist 
in the case of the {\it Core} and 
{\it Mantle} asymmetries
and ii) the statistics in the
{\it Night} sample of Super-Kamiokande 
events is dominated by that 
in the {\it Mantle} sample, 
we expect that
the results obtained using the data on
{\it Core} and {\it Night} asymmetries
without taking into account the
correlation between the two
will not differ substantially
from those derived by 
accounting for the correlation.
The indicated results should not differ much also
from those found by using the data on the 
{\it Core} and {\it Mantle} asymmetries.
Actually, the restrictions following from 
the latter set of data
turned out to be somewhat stronger
that those derived by us on the basis of  
{\it Core} and {\it Night} asymmetry data.

   The results of our study are presented 
graphically in Figs. 1a - 1c and 2a - 2c.
In Figs. 1a - 1c (2a - 2c) the ``conservative'' MSW SMA (LMA and LOW) 
solution region in the $\Delta m^2 - \sin^22\theta$ 
plane of oscillation parameters is shown 
(grey area) together with the regions allowed
by the Super-Kamiokande data on the {\it Core}, {\it Night}
and {\it Mantle} D-N asymmetries at 1.0 s.d. (a), 
1.5 s.d. (b) and 2.0 s.d. (c).
The MSW solution regions were obtained \cite{PK99} 
using the mean event rate solar neutrino data 
\cite{Cl,K,SAGE,GALLEX,SK99}.
Contours of constant {\it Core} D-N asymmetry
in the plane of the two parameters
are also shown (thin solid lines), with the value 
of the asymmetry corresponding to a given contour 
indicated on the contour. The regions allowed by the 
{\it Core}  asymmetry data in Figs. 1a - 1c are 
located to the left  of 
the thick (black) solid line, 
while in Figs. 2a - 2c these regions 
are situated above the upper and below 
the lower thick (black) solid lines.
The regions allowed by the {\it Night}
asymmetry data in Figs. 1a - 1c are 
between the two  dashed 
lines, in Figs. 2a - 2b they are between the 
two upper-most and between the two 
lower-most dashed lines,
and in Fig. 2c they are above 
the upper and below the lower 
dashed lines.
Finally, the regions 
allowed by the {\it Mantle}
asymmetry data in 
Fig. 1a are between the two dash-dotted 
lines, in Figs. 1b - 1c they are 
located to the right of 
the dash-dotted line;
in Figs. 2a - 2c these regions 
are situated 
between the 
two upper-most and between the two 
lower-most dash-dotted lines.
In Figs. 1c and 2c we have shown also the 
contour corresponding to the maximal allowed value
of $A^{C}_{D-N}$ at 3 s.d. 
(the double-thick (black) solid lines).
In obtaining the allowed intervals
of values of 
$A^{N}_{D-N}$, $A^{T}_{D-N}$ and $A^{C}_{D-N}$
at a given C.L. we have added the errors in
eqs. (2), (3) and (4) in quadratures.
For the {\it Core} asymmetry 
this procedure gives the following
1 s.d., 2 s.d. and 3 s.d. allowed intervals 
of values:
$A^{C}_{D-N} = [-0.081,0.046];~[-0.145,0.110];~
[-0.208,0.173]$.
The corresponding intervals of allowed values for the
{\it Night} and {\it Mantle} asymmetries are:
$A^{N}_{D-N} = [0.031,0.099];~[-0.002,0.132];~
[-0.036,0.166]$, and 
$A^{M}_{D-N} = [0.041,0.113];~[0.005,0.149];~
[-0.031,0.185]$.

  Since the minimal value of the predicted {\it Core}
D-N asymmetry in the case of the SMA solution is \cite{SK97II}
$\sim (- 0.03)$ and the lower bound on 
$A^{C}_{D-N}$ following from the current Super-Kamiokande
data at 1 s.d. is $min~(A^{C}_{D-N}) \cong (- 0.081)$,
it is clear that the latter will not play a role 
in constraining the SMA solution region. The current 
Super-Kamiokande upper bound on $A^{C}_{D-N}$, however,
can be used to constrain 
the MSW SMA solution region. 

   As Fig. 1a demonstrates, the MSW SMA ``conservative'' 
solution region is incompatible
with Super-Kamiokande data on the asymmetries 
$A^{N}_{D-N}$ or $A^{T}_{D-N}$
and $A^{C}_{D-N}$, eqs. (2) or (4) and (3), 
including  1 s.d. uncertainties.
The same conclusion is practically valid at 1.5 s.d. as well,
as Fig. 1b shows: the region 
compatible with the 
data on the {\it Night} and {\it Core} asymmetries
in this case is a tiny ``triangle'' around the point
$\Delta m^2 \cong 4\times 10^{-6}~{\rm eV^2}$ and
$\sin^22\theta \cong 0.0085$,
while the {\it Mantle} and {\it Core} asymmetry data
leave allowed only a small point-like
region at $\Delta m^2 \cong 3.9\times 10^{-6}~{\rm eV^2}$ and
$\sin^22\theta \cong 0.0091$.

    At 2 s.d. a large subregion of the SMA ``conservative'' solution
region is ruled out by the Super-Kamiokande data 
on the {\it Core} and {\it Night} D-N asymmetries:
the allowed region in Fig. 1c is located between
the thick dashed line corresponding to 
$min~(A^{N}_{D-N}) = (- 0.002)$
and the dash-dotted line corresponding to
$max~(A^{C}_{D-N}) = 0.110$. 
The allowed region is even smaller if one 
uses the data on the {\it Core} 
and {\it Mantle} asymmetries:
this is the triangular shaped
region to the right of the dash-dotted line 
and to the left of the thick (black) solid line.

   At 3 s.d. level only the data on the {\it Core}
asymmetry constraints the SMA solution region
(Fig. 1c): the subregion located to the right of the
contour corresponding to 
$max~(A^{C}_{D-N}) = 0.173$ 
(double-thick (black) solid line) 
is ruled out by these data.
Most of the indicated subregion 
(not to mention the results described above)
could not be probed at the indicated C.L. 
by the earlier Super-Kamiokande data
on the asymmetry $A^{N5}_{D-N}$ 
associated with the night N5 bin data (see, e.g.,
\cite{SK99,Fogli99,CGG99,PK99}).

   The combined use of the Super-Kamiokande
data on the {\it Core} and {\it Mantle} ({\it Night}) 
D-N asymmetries is less effective in constraining
the MSW LMA solution because the 
{\it mantle-core} enhancement effect 
\cite{SPPLB43498,ChPet991,ChPet992} leads  \cite{SK97II} 
to a modest amplification of the {\it Core}
asymmetry with respect to the {\it Night} 
({\it Mantle}) one in the LMA solution region.
The same conclusion is valid for the LOW solution region.
The results of our analysis for the LMA and LOW solutions
are depicted in Figs. 2a - 2c.
It is interesting that, as Fig. 2a shows,
large parts of
both the LMA and the LOW solution regions
are incompatible at 1 s.d. 
with the Super-Kamiokande data
on the asymmetries $A^{C}_{D-N}$ and $A^{N}_{D-N}$:
the only allowed regions lie in the 
narrow bands corresponding to
$\Delta m^2 \cong (4.0 - 5.0)\times 10^{-5}~{\rm eV^2}$ and
to $\Delta m^2 \cong (1.0 - 2.0)\times 10^{-7}~{\rm eV^2}$.
Even these narrow bands are barely compatible with
the allowed values of $A^{M}_{D-N}$ at 1 s.d.
(Fig. 2a). 

  Similar conclusion is valid if one uses the
D-N asymmetry data, eqs. (2) - (4), 
including the 1.5 s.d. uncertainties, as Fig. 2b shows:
the allowed values of 
$A^{C}_{D-N}$, $A^{M}_{D-N}$ and
$A^{N}_{D-N}$
are compatible 
with the  subregions
located between the upper-most dash-dotted 
and the thick solid lines,
and between the lower-most dash-dotted and 
the thick solid lines. 
The allowed values of $\Delta m^2$ 
in the case of the LMA solution
are confined to the interval
$\Delta m^2 \cong (2.2 - 6.5)\times 10^{-5}~{\rm eV^2}$. 
The region of the LOW solution
corresponding to 
$\Delta m^2 \lesssim 9.5\times 10^{-8}~{\rm eV^2}$ 
is incompatible at the indicated C.L. by the 
current Super-Kamiokande data on the asymmetry
$A^{M}_{D-N}$.

 These data are considerably less restrictive 
at 2 s.d. (Fig. 2c):
the regions between the upper-most
dash-dotted and  thick solid lines,  
and between the lower-most  
dash-dotted and thick solid lines   
are compatible with the data.
The data on the {\it Core} asymmetry alone
including the 3 s.d. uncertainties 
(see Fig. 2c) 
are incompatible 
with the subregion of the LMA ``conservative''
solution region located at 
$\Delta m^2 \lesssim 10^{-5}~{\rm eV^2}$.  


  The solar neutrino data can also be explained assuming
that the solar neutrinos undergo MSW transitions into
sterile neutrino in the Sun: $\nu_e \rightarrow \nu_s$ (see, e.g,
\cite{KLPSter96,PK99}). In this case only a SMA solution
is compatible with the data. 
The corresponding 
solution region obtained (at a given C.L.) 
using the mean event rate solar neutrino data 
and the predictions of
ref. \cite{BP98} for the different 
solar neutrino flux components
practically coincides in  magnitude and shape 
with the SMA $\nu_e \rightarrow \nu_{\mu (\tau)}$
solution region, but is shifted 
by a factor of $\sim 1.2$
along the $\Delta m^2$ axis to smaller 
values of $\Delta m^2$.
The ``conservative'' $\nu_e \rightarrow \nu_s$ 
transition solution region extends both 
in the direction of smaller and larger 
values of $\sin^22\theta$ down to
$0.7\times 10^{-3}$ and up to 0.4~ \cite{PK99}.
The main contribution to the energy-integrated
D-N asymmetries in the three
Super-Kamiokande event samples of interest 
comes from the ``high'' energy tail
of the $^{8}$B neutrino spectrum \cite{SK98III}.
This is related to the fact the relevant 
neutrino effective potential difference
in the Earth matter in the case of 
the $\nu_e \rightarrow \nu_{s}$ transitions
is approximately by a factor of 2 smaller 
than in the case of 
$\nu_e \rightarrow \nu_{\mu(\tau)}$
transitions \cite{Lang87,SK98III}. 
As a consequence, the predicted 
{\it Core} and {\it Night} ({\it Mantle})
D-N asymmetries for the Super-Kamiokande
detector are considerably smaller
in the case of the MSW 
$\nu_e \rightarrow \nu_{s}$ transition
solution than in the case of the 
MSW SMA $\nu_e \rightarrow \nu_{\mu(\tau)}$ solution. 
Nevertheless,
the current Super-Kamiokande data
on the {\it Core} asymmetry
$A^{C}_{D-N}$ including the     
2 s.d. uncertainties, 
as can be shown,
excludes the subregion of the MSW 
$\nu_e \rightarrow \nu_{s}$ 
``conservative'' solution region, located, depending on 
$\Delta m^2$,
approximately at $\sin^22\theta \gtrsim (0.016 - 0.020)$
(see Fig. 6 in \cite{SK98III}).

\vskip 0.3cm
\leftline{\bf III. Conclusions}
\vskip 0.3cm

  We have studied the implications of the
current Super-Kamiokande data on the
{\it Core}, {\it Night} and {\it Mantle} 
D-N asymmetries, eqs. (2) and (3),
for the MSW solutions of the solar neutrino problem.
The Super-Kamiokande collaboration published recently
for the first time data on the {\it Core} asymmetry 
\cite{SK00}. Performing a very simplified analysis 
we have found that practically the whole 
``conservative'' region of
the MSW SMA 
$\nu_e \rightarrow \nu_{\mu(\tau)}$ solution 
in the $\Delta m^2 - \sin^22\theta$ plane
is incompatible with the indicated data 
if one includes the 1 s.d. and 1.5 s.d. uncertainties, 
the only exception in the latter case being 
a point-like
region at $\Delta m^2 \cong 3.9\times 10^{-6}~{\rm eV^2}$ and
$\sin^22\theta \cong 0.0091$.
At 2 s.d. a large subregion of the 
MSW SMA $\nu_e \rightarrow \nu_{\mu(\tau)}$ solution 
region is incompatible with the 
D-N effect data, while at 3 s.d. 
the data on the {\it Core} D-N asymmetry 
{\it alone} excludes a
non-negligible subregion of the 
indicated solution region (Fig. 1c).

   The constraints on the LMA and LOW 
solution regions
from the Super-Kamiokande data on the
{\it Core} and {\it Night} or {\it Mantle} 
D-N asymmetries
are somewhat weaker. Nevertheless,
both the LMA and the LOW solutions
are barely compatible with the 
{\it Core} and {\it Night} asymmetry
data at 1 s.d.: the subregions allowed
by the indicated data
are contained in the narrow bands 
determined by
$\Delta m^2 \cong (4.0 - 5.0)\times 10^{-5}~{\rm eV^2}$ and
 $\Delta m^2 \cong (1.0 - 2.0)\times 10^{-7}~{\rm eV^2}$
(Fig. 2a).
At 1.5 s.d. these data
are incompatible with substantial
subregions
of both solution regions (Fig. 2b), while
at 2 s.d. the LMA solution region 
at $\Delta m^2 \gtrsim 2.0\times 10^{-5}~{\rm eV^2}$
and the whole LOW solution region are allowed 
by the Super-Kamiokande 
{\it Core} and {\it Night} or {\it Mantle}
D-N asymmetry data (Fig. 2c).

  The current Super-Kamiokande data
on the asymmetries
$A^{C}_{D-N}$ and $A^{N}_{D-N}$ allows to 
constrain the MSW 
$\nu_e \rightarrow \nu_{s}$ ``conservative'' 
solution region as well: depending on 
$\Delta m^2$ 
the subregion located approximately at 
$\sin^22\theta \gtrsim (0.016 - 0.020)$
is ruled out at 2 s.d. by the data.

  The simplified analysis  
we have performed demonstrates, in particular, 
the remarkable potential   
which the data on the {\it Night} or {\it Mantle} and especially on 
the {\it Core}  D-N asymmetry have for testing
the MSW $\nu_e \rightarrow \nu_{\mu(\tau)}$ solutions 
of the solar neutrino problem.
As was suggested in \cite{SK97II},
these data are particularly
effective in testing the 
MSW SMA $\nu_e \rightarrow \nu_{\mu(\tau)}$ 
solution. The {\it Core}  D-N asymmetry 
is strongly enhanced in the case of 
MSW SMA $\nu_e \rightarrow \nu_{\mu(\tau)}$ 
solution by the 
{\it mantle-core enhancement} effect
\cite{SPPLB43498,ChPet991,ChPet992}.
With the increase of the statistics
of the Super-Kamiokande experiment 
the data on the two indicated D-N asymmetries
will allow to probe larger and larger 
subregions of the SMA solution region.
Our results indicate that
the data on the {\it Core} and {\it Night}
or {\it Mantle}
asymmetries can be used to 
perform rather effective tests 
of the LMA and LOW 
$\nu_e \rightarrow \nu_{\mu(\tau)}$ solutions 
as well. Using these data one can 
probe and constrain also the MSW 
$\nu_e \rightarrow \nu_{s}$ ``conservative'' 
solution region. Similarly, one can use
the data on the 
{\it Core}, {\it Night} and {\it Mantle} 
D-N asymmetries 
in the charged current event rate 
in the SNO detector to perform equally
effective tests the MSW solutions 
of the solar neutrino problem \cite{MPSNO00}.

\vskip 0.3cm
\leftline{\bf Acknowledgements.}
\vskip 0.3cm 
We would like to thank Y. Suzuki and M. Smy 
for very useful and clarifying 
correspondence concerning the 
Super-Kamiokande data on the D-N effect. 
The work of (S.T.P.) was supported in part by 
the EC grant ERBFMRX CT96 0090.


\leftline{{\bf FIGURE CAPTIONS}}

\vskip 0.3cm
\noindent {\bf Figs. 1a - 1c.}
Constraints on the MSW SMA 
$\nu_e \rightarrow \nu_{\mu(\tau)}$ solution
from the Super-Kamiokande data on the 
{\it Core}, {\it Night} and {\it Mantle} D-N asymmetries,
$A^{C}_{D-N}$, $A^{M}_{D-N}$
and $A^{N}_{D-N}$, eqs. (3), (4) and (2).
The ``conservative'' 
SMA solution region \cite{PK99}
is shown in grey color.
The thick solid line (the (left) 
dash-dotted line) 
corresponds to 
$max~(A^{C}_{D-N})$ ($min~(A^{M}_{D-N})$), 
the dashed lines correspond to
the $min~(A^{N}_{D-N})$ and
$max~(A^{N}_{D-N})$,
allowed by the
Super-Kamiokande data  
at 1 s.d. (a), 1.5 s.d. (b) and 2 s.d. (c).
The double-thick solid line 
in Fig. 1c corresponds to the
3 s.d. maximal value of $A^{C}_{D-N}$
allowed by the data.
Contours of given {\it constant}
{\it Core} D-N asymmetry 
are also shown (thin solid lines).
The region to the left 
of the thick solid line 
(double-thick solid line in Fig. 1c) 
is allowed by the
data on $A^{C}_{D-N}$ (at 3 s.d.),
the region between the two dashed lines
is allowed by the data
on $A^{N}_{D-N}$, while the region
between the two dash-dotted lines (a)
or to the left of the dash-dotted line
(b,c) is allowed by the data on 
$A^{M}_{D-N}$.

\vskip 0.3cm
\noindent {\bf Figs. 2a - 2c.} 
The same as in figures 1a - 1c for the 
MSW LMA and LOW
$\nu_e \rightarrow \nu_{\mu(\tau)}$ 
transition solutions.
The regions allowed by the 
{\it Core} asymmetry data (at 3 s.d.) are
located above the upper and below 
the lower thick solid lines (double-thick solid lines in Fig. 2c).
The regions allowed by the {\it Mantle}
asymmetry data are in Figs. 2a - 2c
between the two upper and between the two 
lower dash-dotted lines,
while those allowed by the {\it Night}
asymmetry data are in Figs. 2a - 2b 
between the two upper and between the two 
lower dashed lines,
and in Fig. 2c they are above 
the upper and below the lower 
dashed lines.
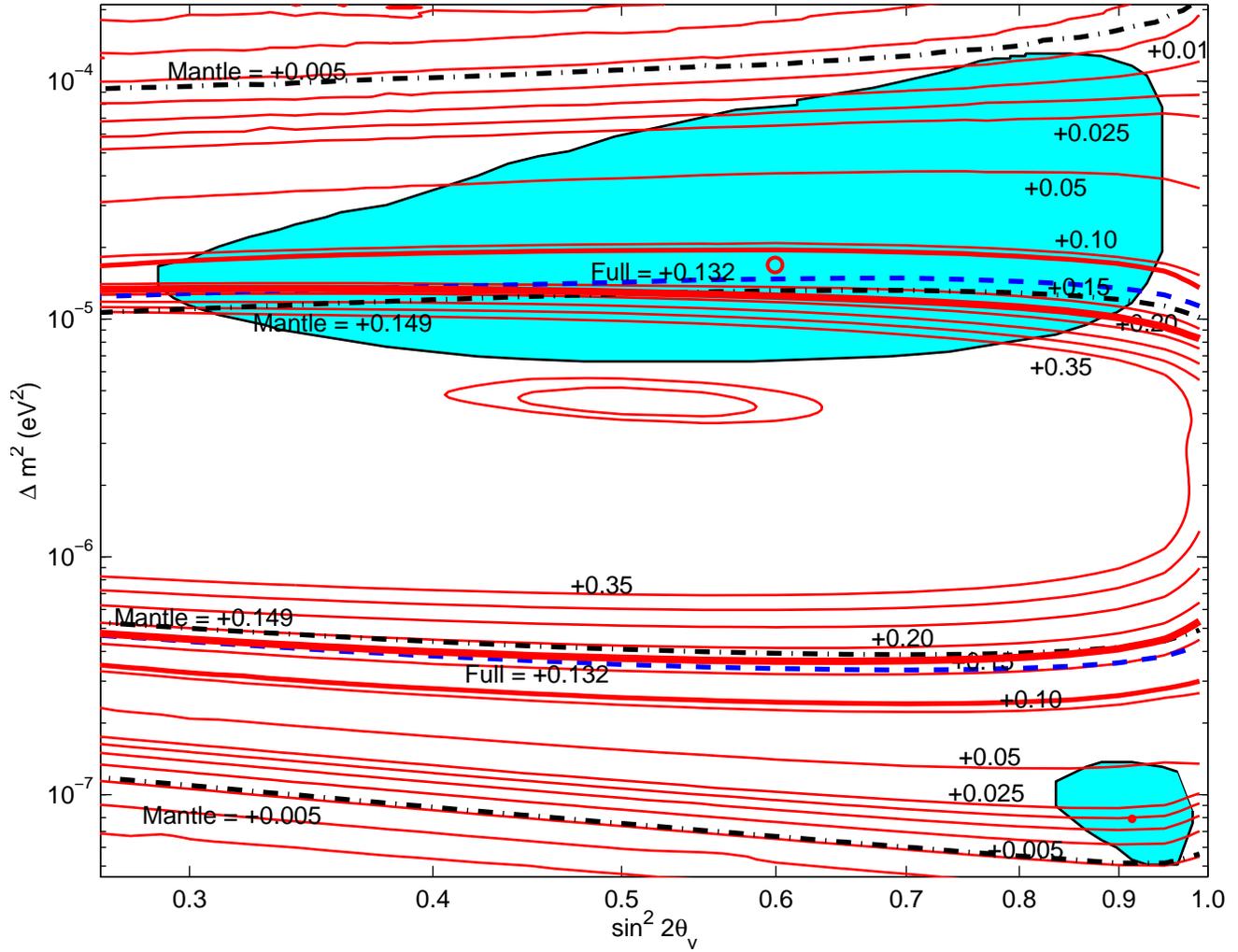
\end{document}